\documentclass[a4,11pt]{article}
\usepackage[margin=2.5cm]{geometry}
\usepackage[affil-it]{authblk}
\usepackage[hidelinks]{hyperref}
\usepackage{graphicx}
\usepackage{amsmath}

\newcommand{\elabel}[1]{\label{eq:#1}}
\newcommand{\Eref}[1]{Eq.~(\ref{eq:#1})}

\newcommand{\flabel}[1]{\label{fig:#1}}
\newcommand{\Fref}[1]{Fig.~\ref{fig:#1}}

\newcommand{\Slabel}[1]{\label{sec:#1}}
\newcommand{\Sref}[1]{Sec.~\ref{sec:#1}}

\title{\textbf{Urban Landscape from the Structure of Road Network: A Complexity Perspective}}

\author[1,*]{Hoai Nguyen Huynh}

\author[]{Muhamad Azfar Bin Ramli}

\affil{Institute of High Performance Computing, Agency for Science, Technology and Research, Singapore}

\affil[*]{\textnormal{\small\textbf{Corresponding author:} \texttt{huynhhn@ihpc.a-star.edu.sg}}}

\date{}

\begin{document}

\maketitle

\abstract{Spatial road networks have been widely employed to model the structure and connectivity of cities. In such representation, the question of spatial scale of the entities in the network, i.e. what its nodes and edges actually embody in reality, is of particular importance so that redundant information can be identified and eliminated to provide an improved understanding of city structure. To address this, we investigate in this work the relationship between the spatial scale of the modelled network entities against the amount of useful information contained within it. We employ an entropy measure from complexity science and information theory to quantify the amount of information residing in each presentation of the network subject to the spatial scale and show that it peaks at some intermediate scale. The resulting network presentation would allow us to have direct intuition over the hierarchical structure of the urban organisation, which is otherwise not immediately available from the traditional simple road network presentation. We demonstrate our methodology on the Singapore road network and find the critical spatial scale to be 85 m, at which the network obtained corresponds very well to the planning boundaries used by the local urban planners, revealing the essential urban connectivity structure of the city. Furthermore, the complexity measure is also capable of informing the secondary transitions that correspond well to higher-level hierarchical structures associated with larger-scale urban planning boundaries in Singapore.}

\vspace{0.2cm}\textbf{Keywords:} Road network, Urban percolation, Spatial organisation, Network presentation, Complexity measure

\section{Introduction}

Characterisation of the existing urban landscape is an important step in helping researchers and planners understand how effective land usage and urban developments have been conducted. While the central question of traditional spatial planning has usually been on the fraction of land to be taken for urban development, it is also important to comprehend the resulting spatial pattern of the built-up areas to understand how these areas are organised or interact with one another. In this respect, the notion of connectivity between places becomes highly important because it strongly relates to the ability of people to move and perform their activities effectively within the urban space, arguably amounting to the functions of different areas. Various approaches have been proposed to study the connectivity pattern of the landscape in both urban and general ecological contexts, including network models \cite{2000@Bunn.etal,2013@Bergsten.Zetterberg} or percolation \cite{2007@Riitters.etal,2009@Bitner.etal,2016@Fluschnik.etal,2019@Behnisch.etal,2021@Montero.etal}, which have contributed to our understanding of the landscape at different scales. Several other studies have also investigated the hierarchical structure of urban systems, including the identification of organic urban boundaries using different methods and data sources, for example, percolation of road junctions \cite{2016@Arcaute.etal}, remote sensing data \cite{2019@Luqman.etal,2020@Cao.etal} or application of scaling laws using a combination of geospatial and census data \cite{2020@Alvioli}.

In any attempt to understand the landscape connectivity of an urban system, the road network is of essential interest as it forms the backbone of any urban system, or city, built upon it \cite{2018@Huynh.etal}. In that role, the network determines the spatial organisation of the system and mediates the socio-economic flows within the city, such as the travel behaviour of people \cite{2012@Parthasarathi.etal,2014@Zhao.etal,2015@Parthasarathi.etal}. The network, therefore, plays an essential role in shaping the urban landscape that would be perceived from different perspectives. While a road network more obviously reflects the results of spatial planning put to an urban system \cite{2013@Barthelemy.etal}, it also exerts a significant impact on the social landscape of the associated system \cite{2018@Gerritse.Arribas-Bel,2019@Zeng.etal}.

Research on urban road networks has been an active subject in various communities, including physics, geography, and information systems. A common theme is the use of spatial networks to represent road systems as a network of nodes (typically representing road junctions) and edges (road segments connecting junctions), which has attracted significant interest in the network literature \cite{2005@Marshall,2011@Barthelemy}. Various related studies include structural quantification of the network, measuring its resilience subjected to fragmentation, as well as providing a proxy to modelling city growth \cite{2004@Jiang.Claramunt,2007@Xie.Levinson,2009@Xie.Levinson,2012@Levinson,2013@Gudmundsson.Mohajeri}. Many measures have also been proposed to compare different road networks in terms of both their topological structure and implications for the corresponding urban development \cite{2004@Jiang.Claramunt,2012@Levinson,2007@Xie.Levinson,2009@Xie.Levinson}.

However, in dealing with road network presentation, it is of fundamental importance to establish what should constitute a node or an edge in such networks. A number of network presentations have been proposed in the literature, including primal \cite{2006a@Porta.etal}, dual \cite{2004@Jiang.Claramunt,2006b@Porta.etal} or line \cite{2015@Marshall} presentations. In a more general sense, a node could either be taken to represent a road intersection or junction, a group of junctions or even a town in a city. In this context, the question of the spatial scale that a node should represent is of particular relevance. It was previously reported that different spatial scales would result in both qualitative and quantitative changes in measurements of the pattern of interest \cite{1989@Turner.etal}. We believe that this is highly related to the amount of information of the system being captured, and one of the main objectives of this study is to construct the relationship between the measured amount of such information as a function of the spatial scale involved. It is, therefore, of interest for us to study how the quantity and quality of the information vary at different spatial scales and study at which optimal point the network would be able to retain the greatest proportion of useful information. More importantly, we investigate what the corresponding network would present when mapped against the urban landscape, giving rise to the notion of places.

The remainder of this paper is organised into 4 sections. Firstly, we describe the procedure to obtain a simplified road network via spatial clustering and aggregation in \Sref{network_presentation}. Secondly, in \Sref{network_measure}, we introduce a complexity-based measure of the information contained within each simplified network for a given level of spatial aggregation. Then we present in \Sref{results} our main results on the relationship between the complexity measure against the level of spatial aggregation applied and discuss the change of the structural network in relation to the physical organisation of road network in an actual urban setting. Finally, conclusions are offered in \Sref{conclusions}.

\section{Network presentation of roads and spatial simplification}
\Slabel{network_presentation}

In modelling roads as a spatial network, nodes are often taken to represent the road junctions, and edges are the road segments between the junctions. For simplicity of discussion in this work, we only consider simple and undirected networks, i.e. an edge between a pair of nodes is uniquely defined, and the relationship between them (e.g. flow) is allowed in both directions. For meaningful analysis, we also exclude any isolated node or cluster of nodes that are not connected to the main network, as well as remove edge from a node to itself (self-loop).

In a network, the degree of a node tells us how many connections it has with others. Those connected to a node are called the (nearest) neighbours of that particular node. In the context of road network, these neighbouring nodes represent the locations that a node can have immediate access to. However, it is realised that nodes of degree 2 play no more than the role of transitional points along a road. In essence, such nodes only enable passage between places, which are indicated by nodes of degree other than 2. In other words, while degree-2 nodes provide connectivity between places, their non-degree-2 counterparts provide accessibility to places. Of the non-degree-2 nodes, it should be noted that the ones with degree 1 appear to be direct indication of a place, while the others (of degree larger than 2) mark where the routes can converge or diverge and where the flows could meet and interchange. As a result, we introduce the so-called structural form of a network (or ``structural network'' for short, if the structure is exclusively discussed without reference to the original network), which is constructed by removing the chunks of degree-2 nodes (and associated edges) and replacing them with direct edges between the pairs of non-degree-2 previously connected to the two ends of the chunks (see \Fref{structural_network}). In this manner, we treat the degree-2 nodes as non-essential and not contributing to the principal structure of the network \cite{2020@Huynh.Selvakumar}.

\begin{figure}
\centering
\includegraphics[width=\textwidth]{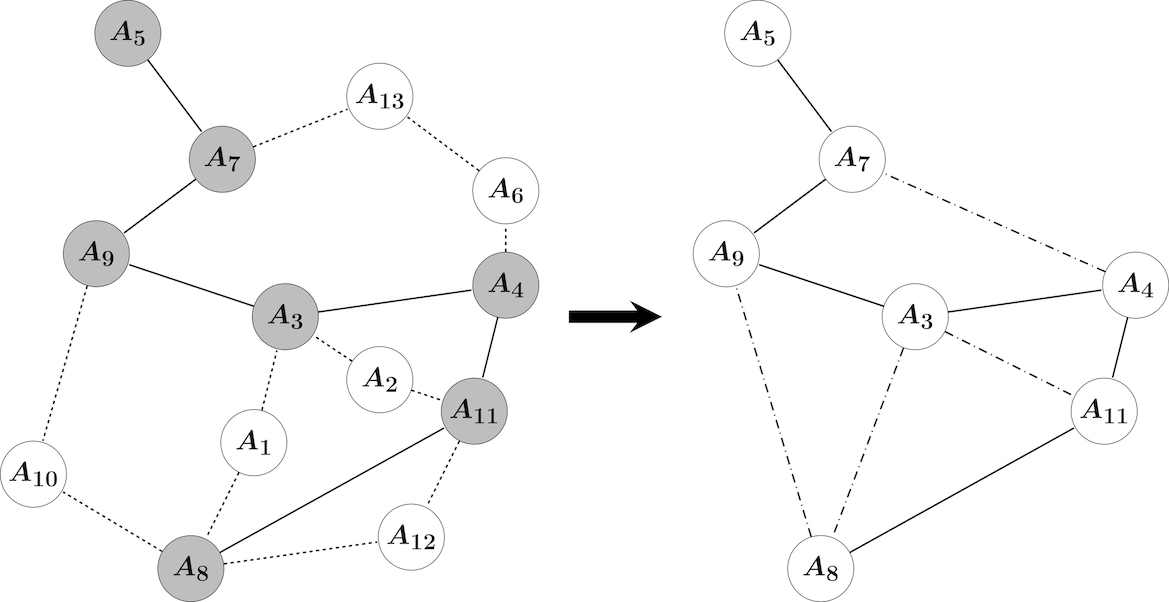}
\caption{\flabel{structural_network}Illustration of structural form of a network. On the left, an example network of 13 nodes is drawn with nodes of degree other than 2 shaded and edges involving degree-2 nodes dashed. On the right, the structural form of the network is extracted by removing all degree-2 nodes and their associated edges, and adding direct edges (dash-dot) between the pairs of corresponding non-degree-2 nodes.}
\end{figure}

When dealing with spatial networks like roads, it is particularly relevant to establish the meaning that each node in the network represents. Typically, a node in such networks represents the physical junction where two or more road segments intersect. However, very often, many nodes, despite their physical distinctions, are redundant in the sense that they are all parts of the same road that provides the connection between the same pair of places. One may argue that it depends on the spatial resolution one is considering to determine whether the generated nodes are actually redundant. Then it really boils down to the question of what scale is appropriate for the representation of a node in a spatial network. After determining that spatial scale, we can find the group of nodes (in the original network) that are (spatially) related and replace them with a representative node in constructing a new (simplified and more abstract) network corresponding to that spatial scale.

Given a spatial scale, the representative nodes in the simplified road network can be found by merging (spatially) nearby nodes in the original network. The merging is done by identifying the cluster of nodes following the procedure in continuum percolation \cite{2019@Huynh} (see also spatial clustering techniques like DBSCAN \cite{1996@Ester.etal}). In such procedure, given a value of the distance parameter $\rho$, we determine the clusters of nodes. A pair of nodes belong to the same cluster if and only if the (Euclidean) distance between them is not larger than the distance threshold $\rho$ (see \Fref{simplified_network}). Each cluster would subsequently be replaced by a representative node located at the centroid of all nodes contained in that cluster.

\begin{figure}
\centering
\includegraphics[width=\textwidth]{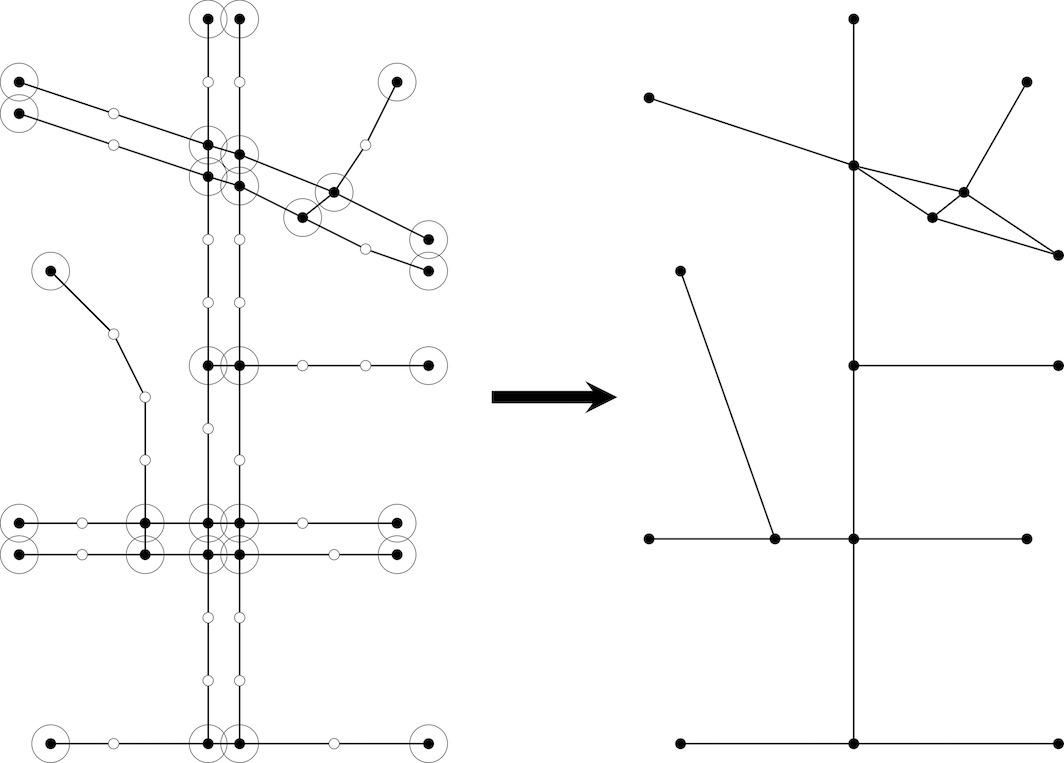}
\caption{\flabel{simplified_network}Simplification of an example road network, both spatially and structurally. Spatial aggregation in a network is performed by clustering non-degree-2 nodes within a certain distance threshold of one another. The structural form of the spatially simplified network is then obtained in the same manner as illustrated in \Fref{structural_network}.}
\end{figure}

\section{Complexity measure of a network}
\Slabel{network_measure}

A network, as a collection of nodes and edges, needs to be quantified by some appropriate measures to characterise the structure of the connections between its nodes and edges. Such measures are important as they can capture the complex structure of the network and allow comparison between different networks. Complexity-based measures prove to be appealing as they encapsulate the mixture of regular and irregular patterns in the system of interest \cite{1989@Crutchfield.Young,2015@Huynh.etal}. Among a number of complexity measures that have been proposed in network literature \cite{2005@Bonchev.Buck,2012@Emmert-Streib.Dehmer,2012@Mowshowitz.Dehmer}, we employ the Shannon entropy formulation for the task in this particular work. In this formulation, the general form for information measure is given by $I=-\sum_i{p_i\log{p_i}}$, in which $p_i$ is the probability of an event $i$  taking place \cite{2018@Zenil.etal}.

Applying this to measure the amount of information stored in the structure of a network, we associate each event with selecting a node carrying a certain property or set of properties. The properties should reflect the topological structure of the network itself, given the connections amongst its nodes. For that purpose, we characterise a node by the probability of being visited by a random walker (or random explorer of the urban road network, also called residing probability, see supplementary material for more info). Using this notion, we measure the complexity of a network by the types of nodes that a random walker would encounter in the network. In this study, the complexity measure $\chi$ of a network is calculated using
\begin{equation}
\elabel{complexity_measure}
\chi = -\sum_\Phi{p_\Phi\log{p_\Phi}}\text{,}
\end{equation}
in which $\Phi$ denotes the label of a node that is potentially visited by the random walker with a certain probability. Some example networks with their corresponding complexity measure value are shown in \Fref{example_networks}.

\begin{figure}
\centering
\includegraphics[width=\textwidth]{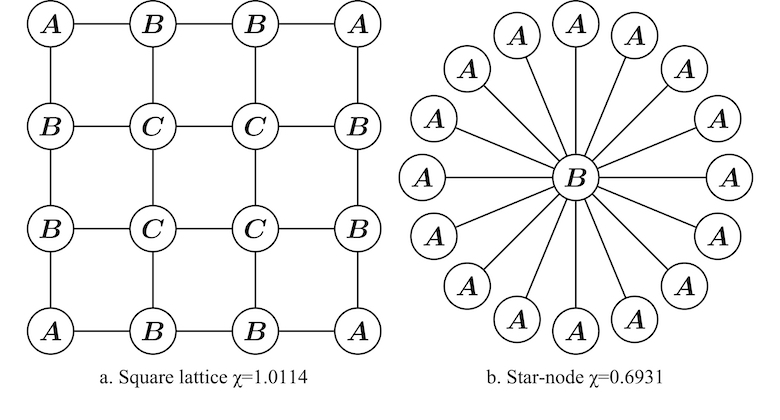}
\caption{\flabel{example_networks}Example networks with different structures and their value of complexity measure as defined in \Eref{complexity_measure}. The nodes with the same letter label share the same residing probability as described in the main text. Corresponding to urban context, a square lattice can be associated with a city with grid-like pattern, a star-node network is similar to a city with every outer part having direct connection with the city centre.}
\end{figure}

\section{Results and discussion}
\Slabel{results}

After introducing the spatial and structural simplification of a network and its calculation of complexity measure, we demonstrate the proposed procedure by applying the analysis to the road network in the city-state of Singapore in order to investigate its spatial structure. The analysis would reveal the pattern of urban organisation in Singapore and how the complexity measure as a measure of pattern mixture varies with different spatial scales imposed on the network. It shall be seen and argued that the complexity measure peaks at some critical value of parameter distance, at which the hierarchical organisation or urban structure would emerge. The structural network is said to experience a phase transition at this critical distance, as it exhibits a stark difference in the form before and after this critical distance.

\subsection{Case study of road network in Singapore}
In order to perform the analysis, we first construct the road network in Singapore using data from OpenStreetMap \cite{OSM}. In this construction, each node in the network literally represents a physical point of zero size in space as it is the intersection between a pair of road segments, which do have any width themselves. As argued earlier, each road junction in reality often contains a number of such road intersections. In order to explore the spatial extent that a node in the network embodies, we define the vicinity of a node in the original network as the area enclosed by a circle of radius $\rho$ centred at the point of interest. As discussed in \Sref{network_presentation}, this parameter $\rho$ would implicitly determine the spatial scale of a node in the simplified network. At every value of the parameter distance $\rho$, the structural network is extracted, and its complexity measure is calculated using \Eref{complexity_measure}.

\begin{figure}
\centering
\includegraphics[width=\textwidth]{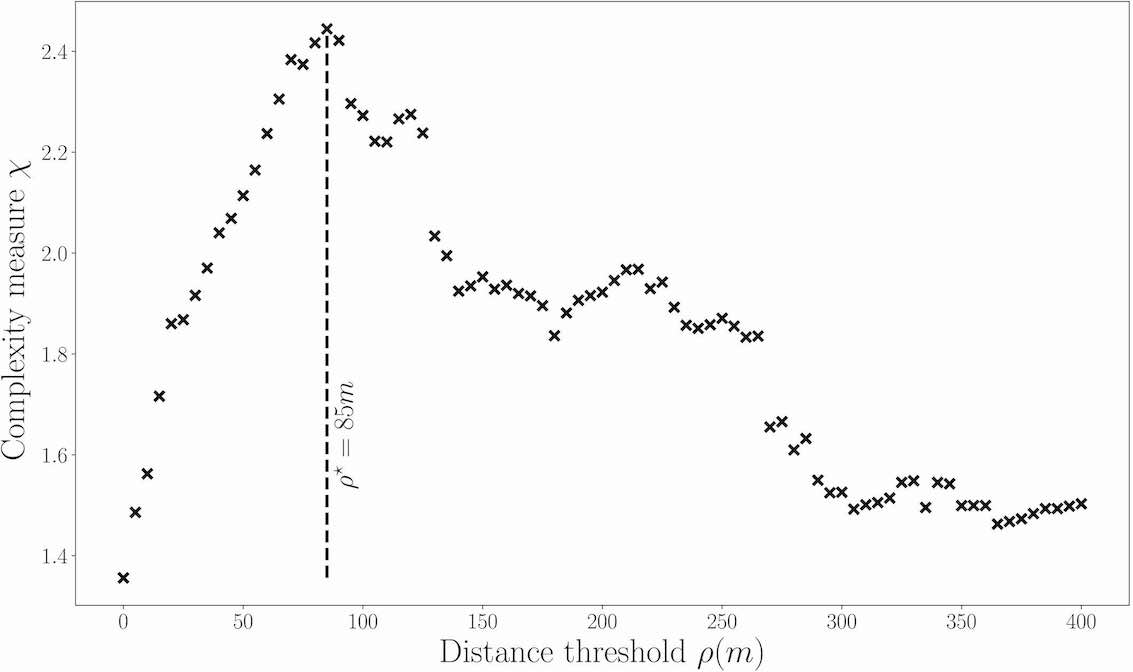}
\caption{\flabel{complexity_curve}Measure of complexity of the structural network at different levels of spatial aggregation of the road network in Singapore. The measure peaks when non-degree-2 nodes in the original road network are clustered at the critical distance threshold of 85 m.}
\end{figure}

\Fref{complexity_curve} shows the complexity measure of the structural network at different levels of spatial aggregation (characterised by parameter distance $\rho$) of the road network in Singapore. At small values of $\rho$, the complexity measure rapidly increases when nodes are absorbed into clusters (which constitute representative nodes in the structural network). This happens as more labels are needed to describe groups of (representative) nodes that now carry more diverse topological properties as discussed in \Sref{network_measure}. After the critical value of $\rho^\star=85$ m, the complexity measure starts to decline as nodes in the original network become over-aggregated, requiring fewer labels to describe fewer groups of diverse topological properties. It is at this critical distance threshold that the structural network reveals the organisational pattern of roads in the original network. At this critical distance, the nodes in the structural network represent the spatial scales below which the detailed connections in the original network are deemed redundant and replaced by the ones in the structural network. In other words, the areas corresponding to these spatial scales are collectively presented by the nodes in the structural network, and the edges among these nodes are the abstract generalisation of connectivity among the areas. Beyond the critical distance threshold, nodes in the original network become over-aggregated, and the spatial scales associated with them are over-presented, being larger than what they could properly encompass.

\subsection{Emergence of town structure and planning areas}

To provide the urban context for the analysis of road network in Singapore, the structural network is overlaid with Master Plan 2014 Planning Area Boundary \cite{Data.gov.sg} to see how the structure matches with the planning as well as the known areas or towns in the city-state. \Fref{structural_network_Singapore} shows the structural network of road connections in Singapore with and without spatial aggregation. It could be observed that without spatial aggregation, the structural network overwhelmingly contains many nodes and edges, which correspond to the actual junctions and connections in the real road network. These nodes and edges, however, make it difficult to comprehend the organisational pattern within the entire urban system. This obscurity is largely due to the fact that most nodes have 3 or 4 connections as they reflect the intersections between roads and that most edges are short, being bounded by the existence of many nodes. In this pattern, different parts of the network appear to be continuously connected with no clear-cut boundary between them. This presentation, on the one hand, is the manifestation of a highly developed and densely constructed urban road network in the city-state of Singapore. On the other, it blurs the demarcation among places and conceals the intrinsic hierarchical structures that are inherently encoded in the organisation of the urban system.

\begin{figure}[h!]
\centering
\includegraphics[width=\textwidth]{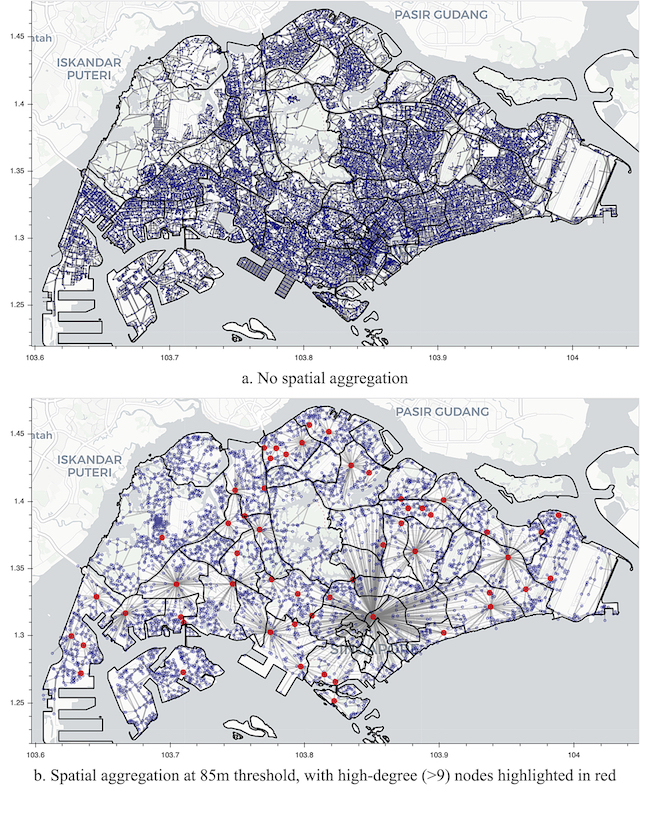}
\caption{\flabel{structural_network_Singapore}Structural network of road connections in Singapore, before and at the critical distance threshold, overlaid with planning boundary.}
\end{figure}

Moving away from this non-aggregated presentation, when nodes in the network are clustered to probe the spatial extent that they appropriately represent, the notion of places starts to enter the picture. As the vicinity of each node is widened, the spatially redundant ones (in terms of the places they represent) become purged. The elimination of this redundancy would reduce the number of noisy edges to give way to more succinct connections among the representative nodes. This operation at the same time also increases the amount of information stored in the abstract connections in the structural network, as seen through the rising complexity measure in \Fref{complexity_curve}. It can be pictured that the complexity measure, in some sense, quantifies the balance of the mixture between the amount of noise removed from the original network and the information accumulated in the representative nodes. At some critical distance threshold where this balance is achieved, places emerge in the form of appropriate clusters of nodes that are spatially related and well delineated. In the structural network, these nodes are characterised by having a high degree as they become the (local) centres of the places they represent and carry radial connections to provide access to the peripheral nodes on the outskirts of the places.

In \Fref{structural_network_Singapore}, the places identified in the procedure above appear to map well to the planning boundary in Singapore. That agreement may not be a surprise as places are designed to be aligned within the planning boundary. Yet, it is noteworthy that the structural and spatial aggregation form of the road network, as a product of abstract presentation of the actual road network, is able to capture the inherent spatial structure that is otherwise not visible in the mere non-simplified network. It should, however, be further noted that not all places map well to the division of planning boundary. In fact, a number of them appear to span over multiple adjacent planning areas. This inevitably suggests that the boundary of the planning areas serves the convenience of planning task and administration by the authority, yet the places in reality may take totally different shapes and organise in different ways which are unintended by the planning.

As an example, we take a closer look at the region in the west of Singapore with notable areas such as Jurong West, Jurong East, Clementi or Choa Chu Kang (see \Fref{structural_network_zoom}) to observe how these places emerge at the critical distance threshold. At a small distance threshold, e.g. 55 m, the structural network still resembles the actual roads, which don't manifest any clear organisational pattern. As the distance threshold increases, roads start to merge in their structural form presentation, and regional centres begin to emerge. At the critical distance threshold of $\rho^\star=85$ m, a clear pattern of organisational structure is revealed, with edges in the structural network forming star-node connections centred at the places mentioned above. Beyond the critical distance threshold, these centres start to merge and absorb into each other (e.g. Jurong West and Jurong East), blurring the boundary between them.

\begin{figure}
\centering
\includegraphics[width=\textwidth]{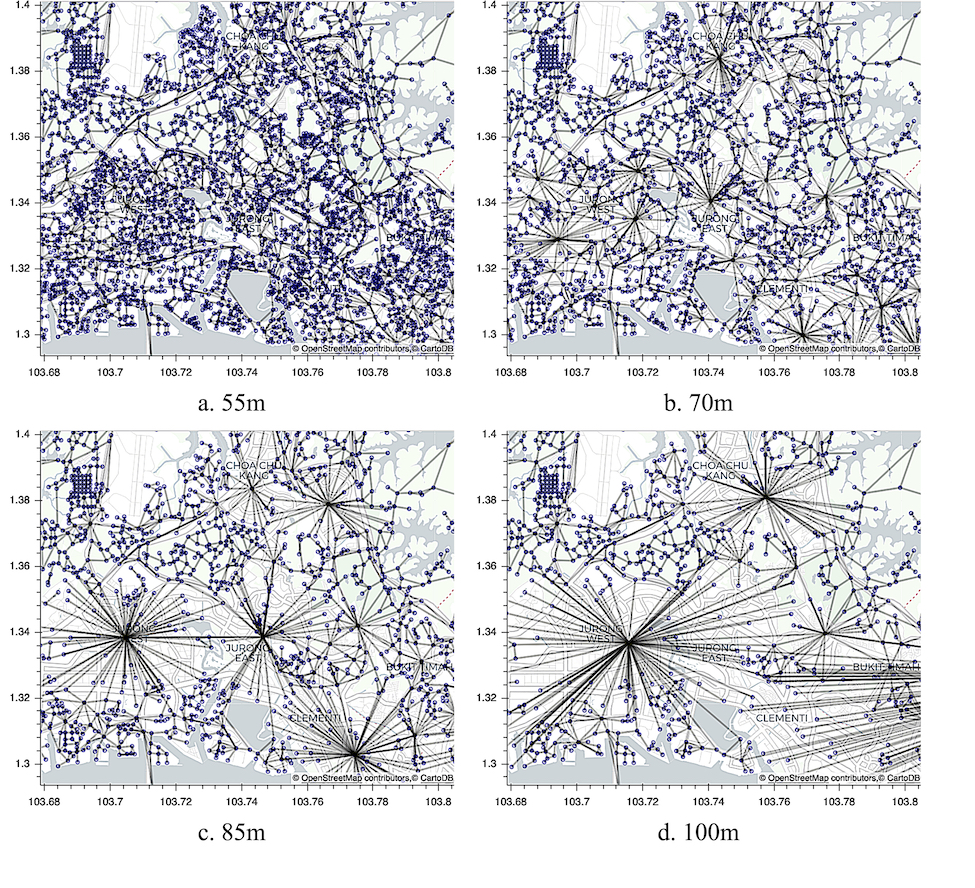}
\caption{\flabel{structural_network_zoom}Structural network of a zoom-in region in western Singapore at different values of distance threshold $\rho$ used for spatial aggregation. A number of areas could be read off the maps, including Choa Chu Kang at the top, Jurong East in the centre, Jurong West on the left, Clementi at the bottom and Bukit Timah on the right, which correspond to some towns and estates in Singapore. A crossover can be observed as $\rho$ varies through the critical value of 85 m when the network transits from a disperse state, where nodes are seen spread, to an over-aggregated state, where nodes and connections are highly centralised to a few places.}
\end{figure}

\subsection{Regional patterns at large spatial scales}

Post critical distance threshold also sees other interesting features of the structural network. Apart from the main peak at $\rho^\star=85$ m, the complexity measure curve in \Fref{complexity_curve} exhibits another two smaller peaks at 120 m and 220 m, respectively. These peaks, though secondary, indicate the change in the spatial organisation pattern in Singapore urban system at larger scales. They indeed correspond to the regional structure that is already well-known in the city-state (see \Fref{structural_network_large_scale}).

\begin{figure}
\centering
\includegraphics[width=\textwidth]{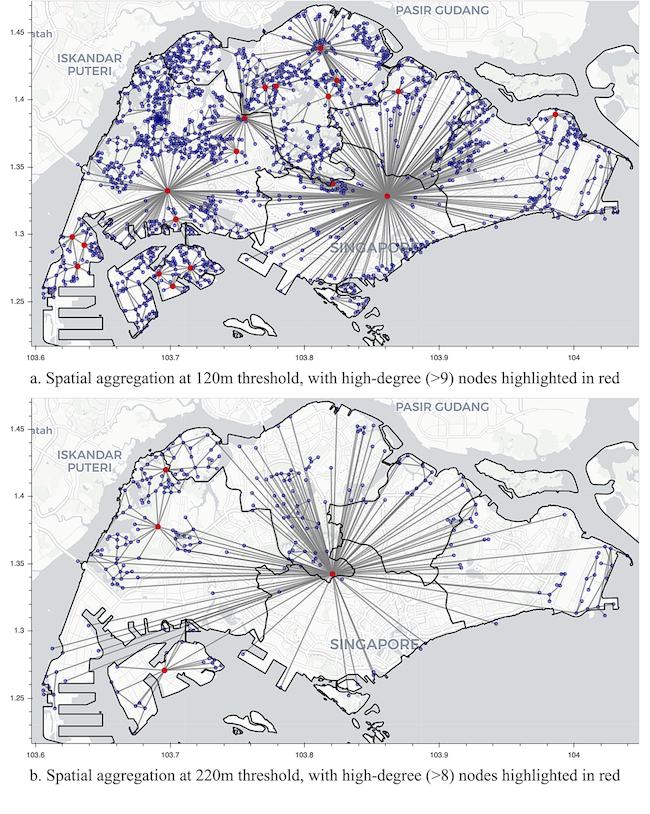}
\caption{\flabel{structural_network_large_scale}Structural network of road connections in Singapore after the critical distance threshold, overlaid with regional boundary. The distance values of 120 m and 220 m are where secondary peaks of complexity measure occur in \Fref{complexity_curve}.}
\end{figure}

At 120 m, the remarkable pattern of the eastern and central regions forming a dominant cluster on the right of the map, alongside the smaller clusters of western (Jurong), north-western (Choa Chu Kang) or northern (Sembawang, Woodlands, Yishun, etc.) regions, emerges as a familiar division of areas in Singapore. It is traditionally known that modern Singapore was developed with root started in the eastern part of the island, where early establishments had already existed in areas such as Marine Parade, Bedok, and Central Area (around Singapore River) or the more recent Toa Payoh. In the last few decades, the focus has been shifted westward with emphasis on the Jurong area (Jurong East and Jurong West). Other major areas in the north have also recently seen more local development, but they remain to be precincts and are still less connected to the rest of Singapore, partly due to the bulk of reserve land and terrain between them and other parts of the country.

Moving to an even larger spatial scale, at 220 m, most of the areas on the main island already congregate to form the largest cluster, which represents the urban part of Singapore. The remaining smaller clusters are the notable remote areas of Western Water Catchment and Lim Chu Kang to the west, and Jurong Island to the south. Largely uninhabitable and located on the fringe of the main island, Western Water Catchment and Lim Chu Kang are mainly planned as areas for military and agricultural purposes, and not included in any residential development plan in the foreseeable future. Jurong Island, on the other hand, is the largest outlying island of Singapore. It was formed by merging several islands in the south-west of the main island through extensive land reclamation efforts of the government. The island is specifically designated to serve the petrochemical industries in Singapore, which have a significant contribution to the economy of the country. Jurong Island, together with Western Water Catchment, are also protected areas with restricted access due to their nature of activities.

\subsection{Interpretation of complexity measure}

Complexity has been a major topic of study across many disciplines. There are many different interpretations of complexity and its measure. In this work, we take the view that complexity is the balance of regularity and irregularity, of order and disorder. In this presentation, the complexity measure is usually plotted against a controlled parameter and shows a hump pattern, in which the complexity measure attains its peak at some intermediate value while remaining low at the two ends \cite{1989@Crutchfield.Young,2019@Huynh,2015@Huynh.etal,2018@Boeing}. In that context, the entropic measure in \Eref{complexity_measure} exhibits this feature with the distance threshold $\rho$ playing the role of a control parameter.

Initially, at a small value of distance threshold, the nodes in the road network are mostly not aggregated, and each of them forms its own cluster. The structural network is then fragmented, with each node having similar and very small residing probability. This makes many nodes share the same label $\Phi$ for having the same residing probability $p_\Phi$, leading to a small value for the complexity measure $\chi$ as computed in \Eref{complexity_measure}. At the other extreme, when the distance threshold is very large, many distant nodes can come together to form a unified cluster, thus reducing the number of nodes as well as the number of labels $\Phi$ needed, effectively curtailing the amount of complexity in the network.

The physical pattern of the structural network also manifests the structural change as the distance threshold varies. It is evident that the network at the critical value of $\rho^\star=85$ m shows striking patterns of towns emerging with connections radial from a centre in a star-node structure. Below this critical value, the network is still pretty much the reflection of the actual road network when many nodes, which correspond to real physical junctions, are present. In such presentation, it is not obvious which nodes play a more central role than the others. Above the critical value of distance threshold, nodes appear to be over-aggregated when places become merged into one another, blurring the boundary between them. The pattern at $\rho^\star=85$ m is where there is a maximal mixture of homogeneity and heterogeneity or the amount of information stored in terms of (structural) network connections. This is exactly at this spatial distance that the town structure emerges, with each town centre being the regional hub for places encompassed within the town.

It should be noted that the curve for complexity measure in \Fref{complexity_curve} shares the same spirit as the results reported in \cite{2017@Molinero.etal}. In that work, the road network is analysed using percolation in terms of the angle formed when road segments meet. However, an important difference between the two approaches is that the goal in \cite{2017@Molinero.etal} is to classify roads in terms of importance and then extract the main skeleton roads. Here in this work, our goal is more on the spatial scale (or spatial embodiment) of nodes in the network and their representation of urban places. Nevertheless, it would be of interest to find out if the two approaches produce comparable results when applying to the same road network, e.g. in the UK. This is, however, beyond the scope of the current study and would be reported elsewhere.

\subsection{Implication for urban planning}

Although there have been different approaches to measure the amount of information contained within a network as well as to quantify its pattern in the urban context, including percolation \cite{2016@Arcaute.etal} or line tessellation model \cite{2017@He.etal}, the complexity measure approach appears to be able to provide deep insight into how the network structure reflects the spatial organisation of the built-up areas in an urban system at various scales. Complex Systems thinking has been applied to view urban systems as systems of many interacting elements or agents, whose overall patterns cannot be fully understood by the local knowledge of each individual part but require a more holistic approach to comprehend its emergent properties and uncover the hidden patterns, which could be translated to appropriate actions in planning and management. The measure of complexity employed in this work encompasses the structure of the network in terms of both the node and edge patterns, and is able to emphasise the mixture of homogeneity and heterogeneity in its spatial pattern. At the lowest level, the nodes embody the spatial units that play similar roles and functions within the urban system. However, the nodes distinguish themselves through the connections with others in the network, giving rise to the heterogeneity in the overall pattern, which amounts to the complexity being captured.

The structurally and spatially simplified network presentation introduced in this work, apart from facilitating the complexity measure calculation, allows the planners to have direct intuition over the hierarchical structure of the urban organisation, which is otherwise not immediately available from the traditional simple road network presentation. This kind of presentation would indeed be useful because it helps to identify the crucial discontinuities in space \cite{2021@Montero.etal}, prompting more attention to those areas in order to ensure effective and integrative usage of land within the urban system. Furthermore, the structural form of the network and its complexity measure could serve as tools to assist the urban planners in understanding the implications of their spatial planning. Beyond urban practitioners to the wider community of urban planning research, the methodology developed in this work can also allow the quantification of the spatial organisation of the built element in functional urban areas. This could then be employed to make a quantitative comparison with other socio-economic or environmental factors \cite{2020@Sapena.etal} to uncover their relationship and help in building simulation models for further study.

\section{Conclusions}
\Slabel{conclusions}
In this work, we investigate the network presentation of actual roads by assessing both the structural and spatial properties of the network. While the structural property refers to the connections among nodes in the network, the spatial one refers to their physical locations in the network's embedding space. We argue that nodes of degree 2 in a network are structurally irrelevant and can be eliminated in constructing the structural form of the network. On the other hand, the spatial scale of the nodes in the network can be probed by defining a distance threshold for the nodes' vicinity and merging the nodes in the same neighbourhood. After that, we quantify the structure of the network by employing an entropic measure of the nodes' residing probability via random walk, which could be interpreted as a complexity measure. Applying this to the case study of the road network in Singapore, we show that the measure stays small at two extreme ends of the distance threshold but peaks at a finite value of the distance. At this critical distance, the structural network shows a pattern in which the organisation of urban structure emerges. The method developed in this work could serve as a systematic procedure to assess landscape connectivity together with its spatial organisation in the urban context. Furthermore, the measure of the spatial structure at different scales can enable the local planning to be put in the perspective of the regional structure of the landscape.

\bibliography{road_complexity}

\end{document}